\documentclass[fleqn,twoside]{article}
\usepackage[headings]{espcrc2}
\usepackage{epsfig}

\readRCS
$Id: espcrc2.tex,v 1.2 2004/02/24 11:22:11 spepping Exp $
\usepackage{graphicx}
\usepackage[figuresright]{rotating}

\newcommand{\AmS}{{\protect\the\textfont2
  A\kern-.1667em\lower.5ex\hbox{M}\kern-.125emS}}

\title{Dynamical parton distribution functions}

\author{Cristian Pisano \address{Vrije Universiteit, \\
 Department of Physics and Astronomy,\\
De Boelelaan 1081, NL-1081 HV Amsterdam, The Netherlands}%
}
       
\runtitle{Dynamical parton distribution functions}
\runauthor{Cristian Pisano}

\begin{document}

\begin{abstract}
Recent measurements for $F_2(x,Q^2)$ have been analyzed
in terms of the `dynamical' and `standard' 
parton model approach at NLO and NNLO
of perturbative QCD. Having fixed the relevant NLO and NNLO parton
distributions, the implications and {\em predictions}
for the longitudinal structure function $F_L(x,Q^2)$ are presented.
It is shown that the previously noted extreme perturbative
NNLO/NLO instability of $F_L(x,Q^2)$
is an artifact of the commonly utilized  `standard' gluon
distributions.  In particular it is demonstrated that using the 
appropriate  -- dynamically generated -- parton distributions at NLO
and NNLO, $F_L(x,Q^2)$ turns out to be perturbatively rather
stable already for $Q^2  \geq {\cal{O}}\, (2-3$ GeV$^2)$.
\vspace{1pc}
\end{abstract}

\maketitle

\section{Introduction}

The parton distributions of the nucleon are determined as a function of the
Bjorken--$x$ variable at a specific low input scale $Q =Q_0$
mainly by experiment, only their evolution to any $Q > Q_0$ being 
predicted by QCD. In the standard framework 
(see, for example, \cite{ref18,ref2}) $Q_0$ is {\em arbitrarily  fixed} at 
some value $Q_0>1$ GeV  and the free parameters of the input 
distributions are varied 
iteratively without any constraint  until the data and the predictions 
yield a minimal $\chi^2$. In this approach even negative gluon distributions  
in the small--$x$ region have been obtained \cite{ref2}, 
leading to
negative 
cross sections like $F_L(x, Q^2)$.  
Alternatively, within the dynamical parton model \cite{grv98,gjr,jr}, the  
distribution functions  at $Q > 1$ GeV are QCD 
radiatively generated from
{\em{valence}}--like (positive) input distributions at an {\em optimally
determined}  $Q_0\equiv \mu < 1$ GeV (where  `valence--like' refers to
$a_f >0$ for {\em{all}} input distributions $xf(x,\mu^2)\sim
x^{a_f}(1-x)^{b_f}$).  This more restrictive ansatz, as compared to the 
standard approach, implies of course less uncertainties \cite{gjr,jr}
concerning the behavior of the parton distributions in the small--$x$ region
at $Q >\mu$, which is entirely due to QCD dynamics at 
$x$ \raisebox{-0.1cm}{$\stackrel{<}{\sim}$} $10^{-2}$.

Following reference \cite{gpr} the 
perturbative stability of the structure function $F_L(x,Q^2)$ in the low 
$Q^2$ region, $Q^2$ \raisebox{-0.1cm}{$\stackrel{<}{\sim}$} 5 GeV$^2$, 
is studied within the framework of the dynamical parton model.
For comparison the same analysis is repeated utilizing a  set 
of `standard' parton distributions previously determined in \cite{ref8}.   

As pointed out in \cite{ref2,ref1,ref3,ref4} the issue of the perturbative 
stability of $F_L(x,Q^2)$ in the very 
small--$x$ region, 
$x$ \raisebox{-0.1cm}{$\stackrel{<}{\sim}$} $10^{-3}$, 
at the perturbatively relevant low values of 
$Q^2$ \raisebox{-0.1cm}{$\stackrel{>}{\sim}$} ${\cal{O}}(2-3$ GeV$^2$), 
is important because it represents a sensitive test of the reliability 
of perturbative QCD.
For the perturbative--order independent rather flat toy model parton
distributions in \cite{ref1}, assumed to be relevant at $Q^2\simeq 2$ 
GeV$^2$, it was shown that next--to--next--to--leading order (NNLO)
effects are quite dramatic at 
$x$  \raisebox{-0.1cm}{$\stackrel{<}{\sim}$} $10^{-3}$
(cf.\ Figure\ 4 of \cite{ref1}).  To some extent such an enhancement
is related to the fact, as will be discussed in more detail in Section 
\ref{sec:fl}, 
that the third--order $\alpha_s^3$ contributions to the longitudinal
coefficient functions behave like $xc_L^{(3)}\sim -\ln x$ at small $x$, as
compared to the small and constant coefficient functions at LO and NLO,
respectively.  It was furthermore pointed out, however, that at higher
values of $Q^2$, say $Q^2 \simeq 30$ GeV$^2$, where the parton
distributions are expected to be steeper in the small--$x$ region
(cf.\ eq.~(13) of \cite{ref1}), the NNLO effects are reduced
considerably.  It is well known that dynamically generated parton
distributions \cite{grv98} are quite steep in the very small--$x$ region
already at rather low $Q^2$, and in fact steeper \cite{gjr} than their
common  `standard' non--dynamical counterparts.  Within this latter
standard approach, a full NLO (2--loop) and NNLO (3--loop) analysis
moreover confirmed \cite{ref2,ref3} the indications for a 
perturbative fixed--order instability observed in \cite{ref1} in the 
low $Q^2$ region. Even additional resummations have been suggested 
\cite{ref7} in order to remedy these instabilities, although such 
additional ad hoc small--$x$ terms lack any quantitative theoretical
basis within QCD.

The plan of this contribution to the proceedings is as follows.
Section 2 is devoted to a description of the formalism used for 
the NNLO and NLO analyses of deep inelastic scattering data within 
the framework 
of the dynamical parton model. In Section 3 the resulting dynamical parton 
distributions of the nucleon and, in particular, their behavior in the 
small--$x$ region are discussed in comparison with their 'standard' 
counterparts.
The predictions for the longitudinal structure function are presented 
in Section \ref{sec:fl}. Summary and conclusions are given in Section 5.

\section{Theoretical framework}

The NNLO and NLO analyses  presented here are performed in the common 
modified minimal 
subtraction  ($\overline{\rm MS}$) factorization and renormalization 
scheme. Heavy quarks ($c$, $b$, $t$) are not considered as massless
partons within the nucleon, i.e. the number of active flavors appearing 
in the splitting functions and the corresponding Wilson coefficients
is taken to be $n_f=3$. This defines the so-called 
'fixed flavor number scheme' (FFNS), which is fully predictive in the heavy 
quark sector: the heavy quark flavors are produced entirely perturbatively
from the initial light ($u$, $d$, $s$) quarks and gluons.
It is 
nevertheless consistent and correct to utilize the standard variable 
$n_f$ scheme for the $\beta$ function \cite{grheavy}.

In the $\overline{\rm MS}$ 
factorization scheme the relevant
 structure function $F_2$ as extracted from the DIS $ep$ process can be,
up to NNLO, written as \cite{curvref3,curvref4,curvref5}
\begin{eqnarray}
\!\!\!\!\!\!\!\!\!\!&&F_2(x,Q^2) =  \nonumber  \\
\!\!\!\!\!\!\!\!\!\!&&F_{2,{\rm NS}}^+(x,Q^2)+ 
F_{2,S}(x,Q^2) + F_2^c(x,Q^2,m_c^2)
\label{eq:f2ms}
\end{eqnarray}
with the non--singlet contribution coming from the three active (light) 
flavors 
being given by 
\begin{eqnarray}
\frac{1}{x}\, F_{2,{\rm NS}}^+(x,Q^2) \!\!\!\! &= \!\!\!\!&
\Big[C_{2,q}^{(0)}+aC_{2,{\rm NS}}^{(1)}+a^2C_{2,{\rm NS}}^{(2)+} 
\Big] \nonumber \\
&&\otimes 
\left[ \frac{1}{18}\, q_8^+ +\frac{1}{6}\, q_3^+\right](x,Q^2)
\end{eqnarray}
where $\otimes$ denotes the common convolution, 
$a=a(Q^2)\equiv\alpha_s(Q^2)/4\pi$, $C_{2,q}^{(0)}(z)=\delta(1-z)$,
$C_{2,{\rm NS}}^{(1)}$ is the common NLO coefficient function (see,
for example, \cite{curvref6}) and a convenient expression for the relevant 
NNLO 2-loop Wilson coefficient $C_{2,{\rm NS}}^{(2)+}$ can be found 
in \cite{curvref3}.
The NNLO $Q^2$-evolution of the flavor 
{\mbox{non-singlet} 
combinations $q_3^+=u+\bar{u}-(d+\bar{d})=u_v-d_v$ and $q_8^+=
u+\bar{u}+d+\bar{d}-2(s+\bar{s}) = u_v+d_v+4\bar{q}-4\bar{s}$, where
$\bar{q}\equiv\bar{u}=\bar{d}$ and $s=\bar{s}$, is related to the 
3-loop splitting function \cite{curvref7} $P_{\rm NS}^{(2)+}$, besides 
the usual LO (1-loop) and NLO (2-loop) ones, $P_{\rm NS}^{(0)}$ and
$P_{\rm NS}^{(1)+}$, respectively \cite{curvref3,ref16}.  Notice that we 
do not consider sea breaking effects ($\bar{u}\neq\bar{d},\,\, s\neq
\bar{s}$) since the HERA data used, and thus our analysis, are not
sensitive to such corrections.  The flavor singlet contribution in
(1) reads 
\begin{eqnarray}
\frac{1}{x}\, F_{2,S}(x,Q^2)\!\!\!\! &=\!\!\!\! & \frac{2}{9}
\left\{ \left[ C_{2,q}^{(0)} + aC_{2,q}^{(1)}+a^2C_{2,q}^{(2)}\right]\right .
\otimes \Sigma \nonumber \\ 
&& \!\!\!\!\!\!\!\!\!\!\!\!+
\left.\left[aC_{2,g}^{(1)}+a^2C_{2,g}^{(2)}\right] \otimes g\right\} (x,Q^2)
\end{eqnarray}
with 
$\Sigma(x,Q^2)\equiv\Sigma_{q=u,d,s}(q+\bar{q})=u_v+d_v+4\bar{q}+2\bar{s}$,
$C_{2,q}^{(1)}=C_{2,\rm NS}^{(1)}$ and the additional common NLO gluonic
coefficient function $C_{2,g}^{(1)}$ can be again found in \cite{curvref6},
for example.  Convenient expressions for the NNLO $C_{2,q}^{(2)}$ and
$C_{2,g}^{(2)}$ have been given in \cite{curvref4} and the relevant 3-loop
splitting functions $P_{ij}^{(2)}$, required for the evolution of 
$\Sigma(x,Q^2)$ and $g(x,Q^2)$, have been derived in \cite{curvref9}.  We
have performed all $Q^2$-evolutions in Mellin $n$-moment space and used
the QCD-PEGASUS program \cite{ref9} for the NNLO evolutions, appropriately 
modified to account for the fixed
$n_f=3$ flavor number scheme with a running $\alpha_s(Q^2)$.  In NNLO
the strong coupling evolves according to 
\begin{equation}
\frac{d a}{ d \ln Q^2} = -\Sigma_{\ell =0}^2\, \beta_{\ell}\, a^{\ell +2}\, , 
\end{equation}
where
$\beta_0 = 11-2f/3$, $\beta_1=102-38f/3$ and $\beta_2=2857/2-5033f/18+
325f^2/54$ and the running $a(Q^2)$ is appropriately matched at 
$Q=m_c$  and $Q=m_b$. The values $m_c= 1.3$ GeV 
and $m_b= 4.2$ GeV have been used, as implied by optimal fits
\cite{gjr} to recent deep inelastic $c$-- and $b$--production HERA
data.

\maketitle
\begin{table*}[ht]
\setlength{\tabcolsep}{1.pc}
\newlength{\digitwidth} \settowidth{\digitwidth}{\rm 0}
\catcode`?=\active \def?{\kern\digitwidth}
\caption{Parameter values of the dynamical NNLO and NLO QCD fits
with the parameters of the input distributions referring to 
(\ref{eq:input}) at a common input scale $Q_0^2=\mu^2=0.5$ GeV$^2$ optimal 
at both perturbative orders. Here $\chi^2$ was evaluated by adding in 
quadrature the statistical and systematic errors. }
\label{tab:dyn}
\centering
\begin{tabular*}{\textwidth}{@{}l@{\extracolsep{\fill}}rrrrrrrrr}
\hline
& \multicolumn{4}{c}{NNLO}  & 
\multicolumn{4}{c}{NLO} \\
\hline
& $u_v$ & $d_v$ & $\bar{q}$ & $g$ &
  $u_v$ & $d_v$ & $\bar{q}$ & $g$ \\

N &  0.6210  &   0.1911     & 0.4393    &  20.281   &
     0.5312   &   0.3055     & 0.4810    &  20.649  \\

a &    0.3326      &  0.8678     & 0.0741    &  0.9737    &
       0.3161      &  0.8688     & 0.0506    &  1.3942       \\ 

b &    2.7254     &  4.7864      &  12.624   &   6.5186   &
       2.8205     &  4.6906      &  14.580   &   11.884       \\

c &    -9.0590     & 65.356        & 2.2121  & ---    &
       -8.6815     & 44.828        & -2.2622 &  15.879    \\

d &  53.547      &   1.6215        & 7.7450    &  ---      &
     54.994      &  -5.3645        &  21.650   &  ---    \\

e &   -36.979    &    -41.117        & --- & --- &
      -40.088    &     -21.839       & --- & --- \\

$\chi^2/{\rm dof}$ & \multicolumn{4}{c}{1.037} & 
                     \multicolumn{4}{c}{1.073} \\
$\alpha_s(M_Z^2)$ & \multicolumn{4}{c}{0.112} & 
                      \multicolumn{4}{c}{0.113} \\

\hline
\end{tabular*}
\end{table*}
\begin{table*}[ht]
\setlength{\tabcolsep}{1.pc}
\catcode`?=\active \def?{\kern\digitwidth}
\caption{As Table \ref{tab:dyn} but for the standard NNLO and NLO QCD 
fits with the parameters of the input distributions referring to 
(\ref{eq:input}) at a common input scale $Q_0^2 = 1.5$ GeV$^2$.} 
\label{tab:std}
\centering
\begin{tabular*}{\textwidth}{@{}l@{\extracolsep{\fill}}rrrrrrrrr}
\hline
& \multicolumn{4}{c}{NNLO}  & 
\multicolumn{4}{c}{NLO} \\
\hline
& $u_v$ & $d_v$ & $\bar{q}$ & $g$ &
  $u_v$ & $d_v$ & $\bar{q}$ & $g$ \\

N & 0.2503 & 3.6204 & 0.1196 & 2.1961 &
    0.4302 & 0.3959 & 0.0546 & 2.3780 \\

a & 0.2518 & 0.9249 & -0.1490 & -0.0121 &
    0.2859 & 0.5375 & -0.2178 & -0.0121 \\ 

b & 3.6287 & 6.7111 & 3.7281 & 6.5144 &
    3.5503 & 5.7967 & 3.3107 & 5.6392 \\

c & 4.7636 & 6.7231 & 0.6210 & 2.0917 &
    1.1120 & 22.495 & 5.3095 & 0.8792 \\

d & 24.180 & -24.238 & -1.1350 & -3.0894 &
    15.611 & -52.702 & -5.9049 & -1.7714 \\

e & 9.0492 & 30.106 & --- & --- &
    4.2409 & 69.763 & --- & --- \\

$\chi^2/{\rm dof}$ & \multicolumn{4}{c}{0.989} & 
                     \multicolumn{4}{c}{0.993} \\
$\alpha_s(M_Z^2)$ & \multicolumn{4}{c}{0.112} & 
                      \multicolumn{4}{c}{0.114} \\

\hline
\end{tabular*}
\end{table*}
  
The heavy flavor
(dominantly charm) contribution $F_2^c$ in (1) 
is taken as in \cite{ref8,curvref2} as given
by the fixed-order NLO perturbation theory \cite{ref10,ref11}.  The small 
bottom contribution turns out to be negligible.
These contributions are gluon $g(x, \mu_F^2)$ dominated where the
factorization scale should preferably be chosen \cite{Gluck:1993dpa} 
 to be $\mu_F^2= 4 m_h^2$. 
It has been shown \cite{gjr} that the resulting predictions are in 
perfect agreement with all available DIS data on heavy quark production
and are furthermore perturbatively stable \cite{Gluck:1993dpa}.
Even choosing a very large scale like $\mu_F^2 = 4 (Q^2 +  4 m_c^2)$
leaves the NLO results essentially unchanged \cite{Gluck:1994uf,Vogt:1996wr},
 in 
particular in the small--$x$ region. 
This stability renders attempts to resum supposedly
'large logarithms' ($\ln Q^2/m_h^2$) in heavy quark production cross
sections superfluous. Since a NNLO calculation of heavy quark production is
 not yet
available, we have again used the same NLO ${\cal{O}}(\alpha_s^2)$
result.  This is also common in the literature \cite{ref12,ref13,ref14}
and the error in the resulting parton distributions due to NNLO 
corrections to heavy quark production is expected \cite{ref12} to be 
less than their experimental errors. 
 
\section{Quantitative results}

For the present analysis the valence--like
input distributions at $Q_0\equiv \mu < 1$ GeV are parametrized according
to \cite{gpr,ref8,curvref2}
\begin{eqnarray}
xq_v(x,Q_0^2) \!\!\!\! & = \!\!\!\!& N_{q_v}x^{a_{q_v}}(1-x)^{b_{q_v}}
      (1+c_{q_v}\sqrt{x} \nonumber \\
&&+d_{q_v}x +e_{q_v}x^{1.5})\, ,\nonumber \\
xw(x,Q_0^2) \!\!\!\! & =\!\!\!\! & 
N_w x^{a_w}(1-x)^{b_w}(1+c_w\sqrt{x}+d_w x)\, ,
\nonumber \\
\label{eq:input}
\end{eqnarray}
for the valence $q_v=u_v,\, d_v$ and sea $w=\bar{q},\, g$ densities.
Since the data sets utilized are insensitive to the specific choice of
the strange quark distributions, we generate the strange densities
entirely radiatively \cite{grv98}, starting 
from $s(x,Q_0^2)=\bar{s}(x,Q_0^2)=0$, where $Q_0<1$ GeV.
The normalizations $N_{u_v}$ and $N_{d_v}$
are fixed by $\int_0^1 u_v dx = 2$ and $\int_0^1 d_v dx=1$,
respectively, and $N_g$ is fixed via $\int_0^1 x(\Sigma +g)dx=1$.
The following data sets
have been used:  the small-$x$ \cite{ref28} and large-$x$ 
\cite{ref28b} H1 $F_2^p$ data; the fixed target BCDMS data 
\cite{ref28c} for $F_2^p$ and $F_2^n$ using $Q^2\geq 20$ GeV$^2$
and $W^2=Q^2(\frac{1}{x}-1)+m_p^2\geq 10$ GeV$^2$ cuts, and the 
proton and deuteron NMC data \cite{ref28d} for $Q^2\geq 4$ GeV$^2$
and $W^2\geq 10$ GeV$^2$.  This amounts to a total of 740 data 
points.
 The required overall 
normalization factors of the data turned out to be 0.98 for H1
and BCDMS, and 1.0 for NMC. We use here solely deep
inelastic scattering data since we are mainly interested in the
small--$x$ behavior of structure functions. 
 The resulting parameters
of the NLO and NNLO fits are summarized in Table \ref{tab:dyn}.
The corresponding dynamical gluon and sea distributions, 
evolved to some specific values 
of $Q^2 > Q_0^2$, are  very similar to the ones in 
\cite{gjr,jr} which were obtained from a global analysis including
Tevatron Drell--Yan dimuon production and, at the NLO level, high--$E_T$ 
inclusive jet data as well. Furthermore, it
should be mentioned
that the NLO $\alpha_s(M_Z^2)$ in Table \ref{tab:dyn} 
turns out be somewhat smaller
in fits based solely on deep inelastic structure function data 
\cite{ref8,ref16,ref12,ref15,ref17} as compared to those which take
into account additional hard scattering data \cite{ref18,ref2,gjr,ref19}
(for a recent summary, see \cite{ref20}).  At NNLO the resulting
$\alpha_s(M_Z^2)$ is generally slightly smaller \cite{ref20} (c.f.\
Table \ref{tab:dyn}) which is due to the fact that the higher the perturbative
order the faster $\alpha_s(Q^2)$ increases as $Q^2$ decreases.  

\begin{figure}[htb]
\vspace{-2.5cm}
\hspace{-0.5cm}
\epsfig{figure= 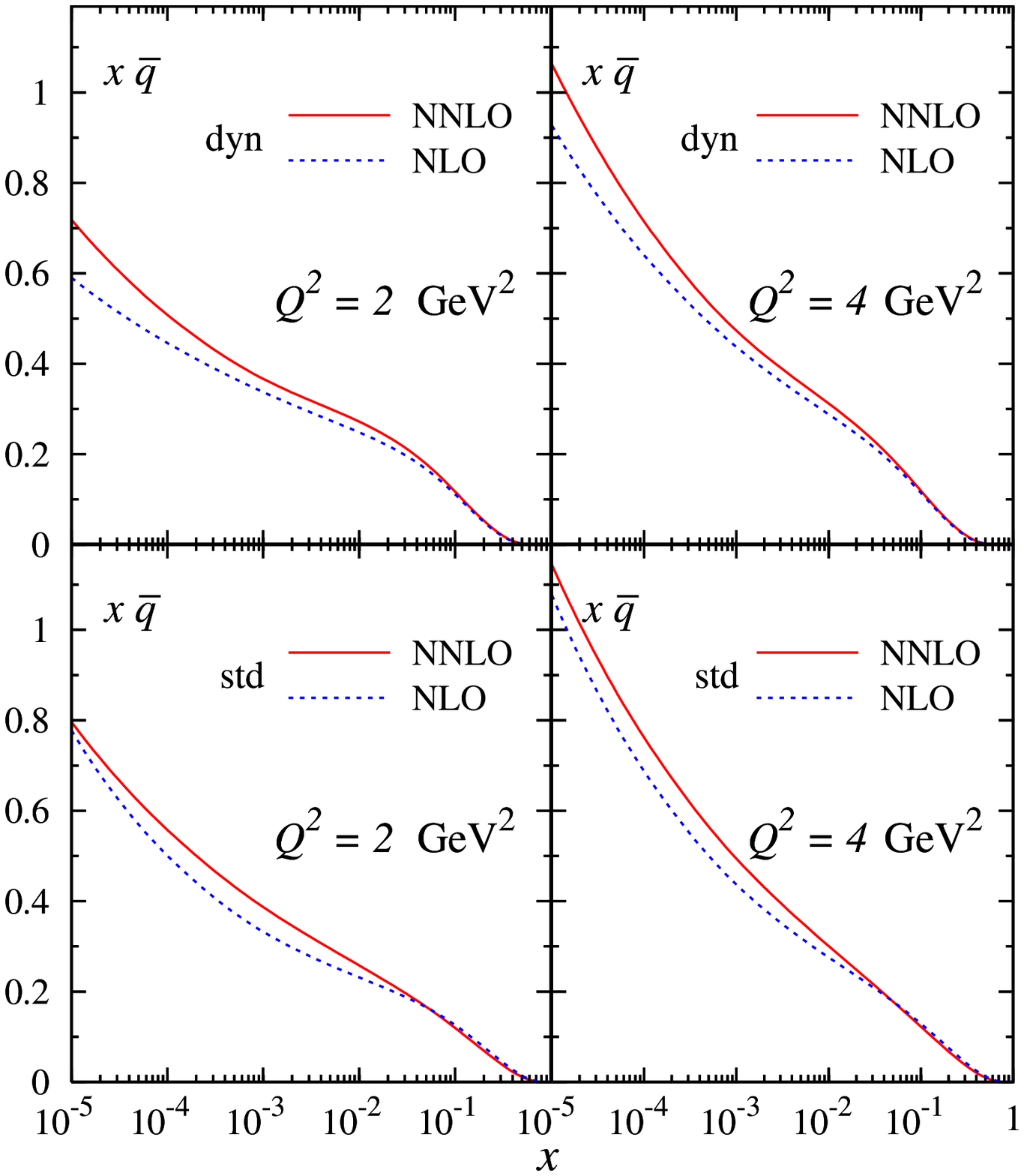, width = 8.cm}
\vspace{-1.5cm}
\caption{The  
sea distributions 
$x \bar q (x, Q^2)$, where $\bar q \equiv \bar u = \bar d$, 
in the dynamical (dyn) and standard (std) parton model 
for two representative low values of $Q^2$.}
\vspace{-0.4cm}
\label{fig:xqs}
\end{figure}
\begin{figure}[htb]
\vspace{-2.5cm}
\hspace{-0.6cm}
\epsfig{figure= 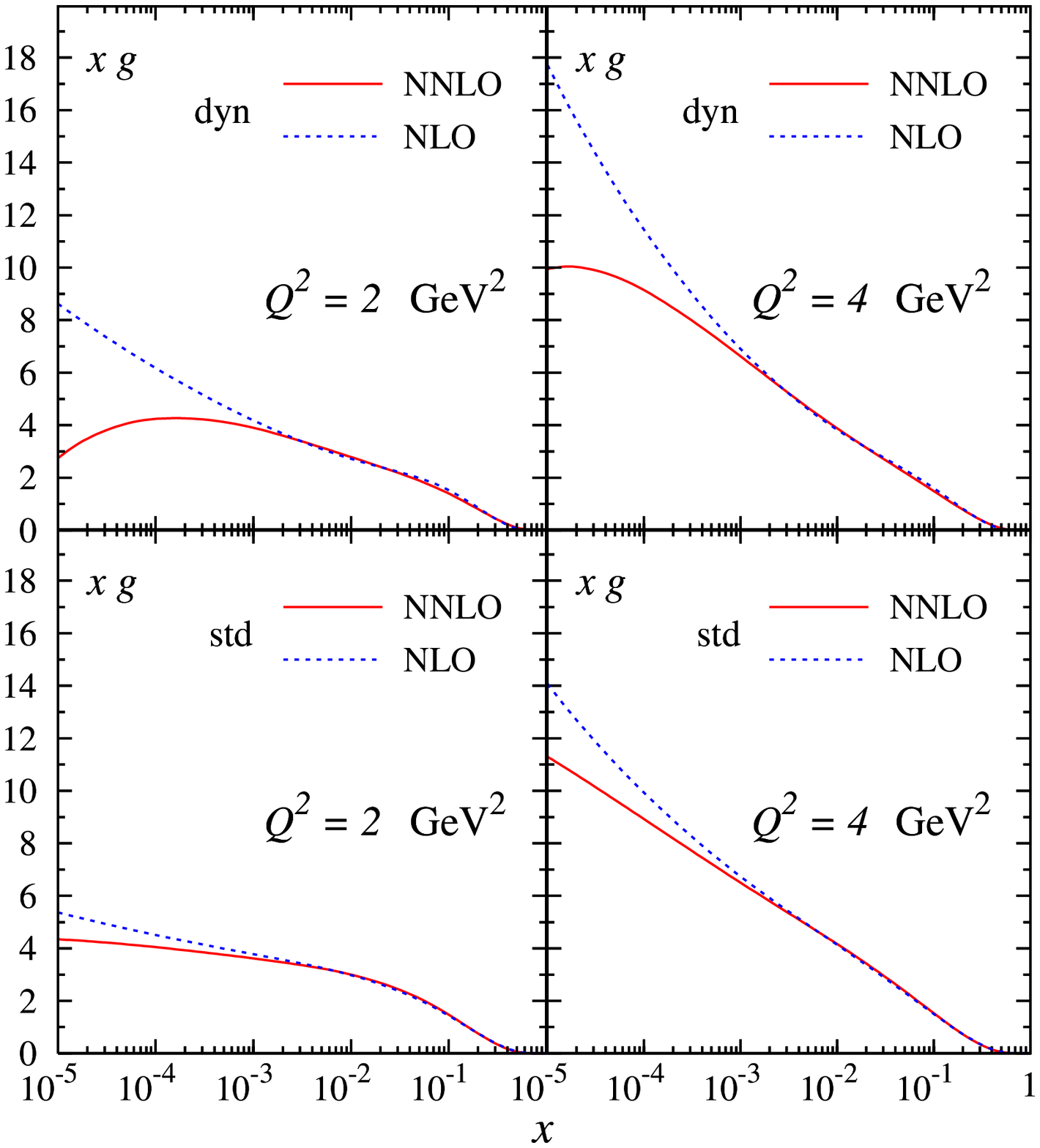, width = 8.cm}
\vspace{-1.5cm}
\caption{As in Figure \ref{fig:xqs}, but for the gluon distribution 
$x g (x, Q^2)$.}
\vspace{-0.4cm}
\label{fig:xg}
\end{figure}

For comparison the results of a 'standard' fit to the same data
\cite{gjr} are presented in Table \ref{tab:std},
 where the gluon and sea input 
distributions in (\ref{eq:input}) do not vanish as $x\to 0$ 
($a_{g,\bar q}$ \raisebox{-0.1cm}{$\stackrel{<}{\sim}$} 0) at $Q_0^2 = 1.5$ 
GeV$^2$. Without loss of generality the 
strange sea at the input scale is taken to be $s(x, Q_0^2)
=\bar{s}(x, Q_0^2) =0.5\, \bar{q}(x, Q_0^2)$.
 The overall 
normalization factors of the data are 0.98 for BCDMS and 1.0 
for H1 and NMC, while slightly different values of the charm and bottom
masses have been used, namely $m_c=1.4$ GeV and $m_b=4.5$ GeV.
The standard NNLO and NLO fits are very similar to each other, 
corresponding to $\chi^2/{\rm dof}=0.989$ and $0.993$, 
respectively, with the  NNLO predictions for $F_2$ falling slightly
below the  NLO  ones at smaller
values of $Q^2$ \cite{ref8}.  
It should be emphasized that the perturbatively
stable QCD predictions, both in the dynamical and standard approaches,
 are in perfect agreement with all recent
high-statistics measurements of the $Q^2$-dependence of 
$F_2(x,Q^2)$ in the (very) small-$x$ region.  Therefore
additional model assumptions concerning further resummations of
subleading small-$x$ logarithms (see, for example, \cite{curvref19})
are not required \cite{curvref7,curvref9}.

The sea and gluon distributions resulting from both fits 
are shown in Figures \ref{fig:xqs} and \ref{fig:xg} respectively. 
The dynamical NLO sea distribution has a rather similar
small--$x$ dependence as the   standard one \cite{gjr,ref8};
this is caused by the fact that
the valence--like sea input in (\ref{eq:input}) 
vanishes very slowly as $x\to 0$
(corresponding to a small value of $a_{\bar{q}}$, $a_{\bar{q}}\simeq
0.05$, according to Table \ref{tab:dyn}) and thus is similarly 
increasing with decreasing $x$ down to $x\simeq 0.01$ as the sea input 
obtained by
a  standard fit. On the other hand, the dynamically generated NLO gluon is 
steeper as $x\to 0$ than the gluon distributions obtained from the
 standard fits. Similar remarks hold when comparing dynamical and
standard distributions at NNLO. At  NNLO the sea distribution $x\bar{q}$ is 
larger (steeper) than the NLO one, whereas the NNLO gluon distribution $xg$
is flatter as $x$ decreases and, in general, falls below the NLO one
in the small--$x$ region. It is evident from Figure 
\ref{fig:xg} that the NNLO gluon remains valencelike even at 
$Q^2 = 2-4$ GeV$^2$, i.e. decreases as $x$ decreases; this is mainly 
caused by the dominant NNLO gluon-gluon splitting function 
$P_{g g}^{(2)}$ which is {\em negative} and more singular as $x\to 0$ than 
the LO and NLO ones, $P_{g g}^{(2)}(x) \sim \frac{1}{x} \ln \frac{1}{x}$
\cite{jr}.

\begin{figure}[htb]
\vspace{-1.6cm}
\hspace{-0.6cm}
\epsfig{figure= 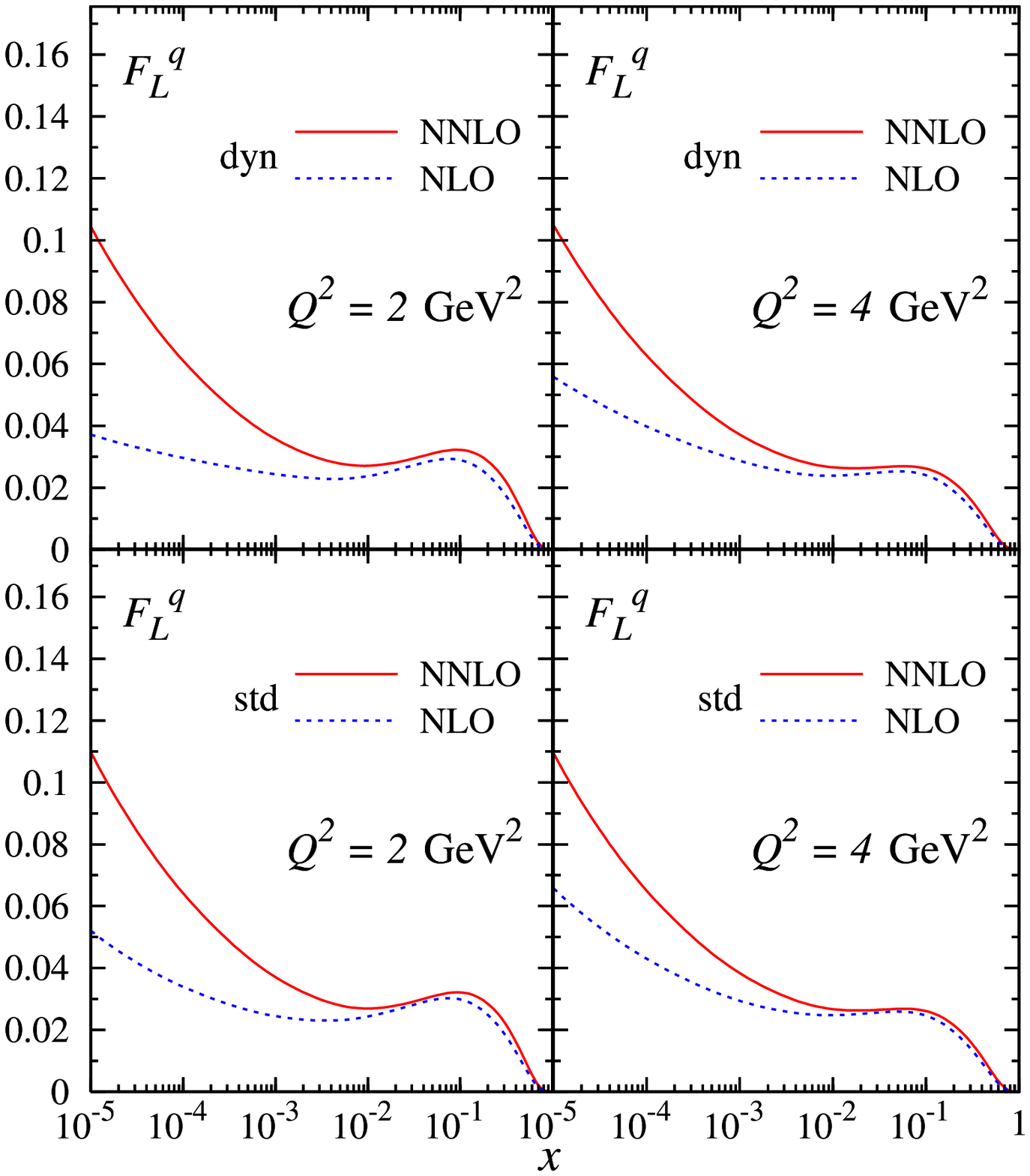, width = 8.cm}
\vspace{-1.5cm}
\caption{The individual light ($u,d,s$) quark contribution $F_L^q$ to
the total $F_L$ in (\ref{eq:flms}) in the dynamical (dyn) and standard  
(std) parton
approach at NNLO and NLO for two representative low values of $Q^2$.
The standard NLO results in the lower panel 
are similar for the CTEQ6 (anti)quark distributions
\cite{ref18}. Notice that, according to (\ref{eq:flms}), 
$F_L^q + F_L^g = F_L - F_L^c$.}
\vspace{-0.4cm}
\label{fig:flq}
\end{figure}
\begin{figure}[htb]
\vspace{-1.6cm}
\hspace{-0.6cm}
\epsfig{figure= 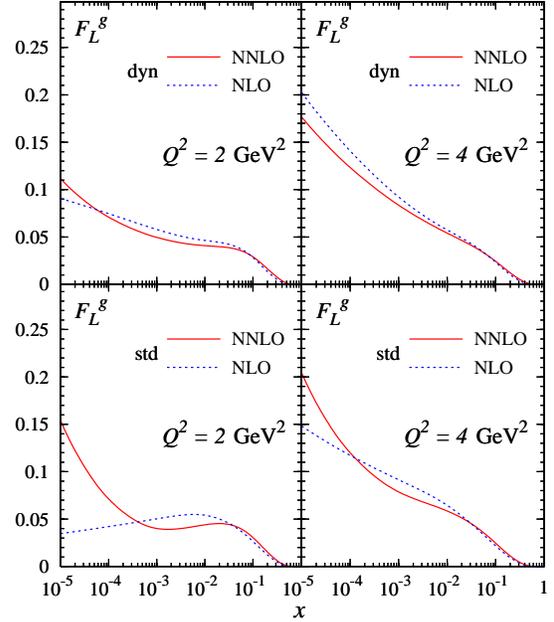, width = 8.cm}
\vspace{-1.5cm}
\caption{As in Figure~\ref{fig:flq} but for the gluonic contribution 
$F_L^g$ to $F_L$
in (\ref{eq:flms}) with $F_L^g=\frac{2}{9}x\, C_{L,g}\otimes g$.}
\vspace{-0.4cm}
\label{fig:flg}
\end{figure}
\begin{figure}[htb]
\vspace{-1.6cm}
\hspace{-0.6cm}
\epsfig{figure= 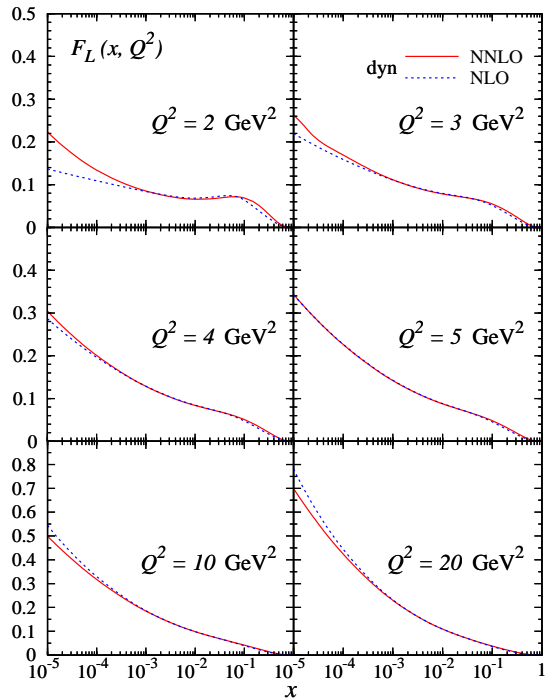, width = 8.cm}
\vspace{-1.5cm}
\caption{Dynamical parton model NNLO and NLO predictions for $F_L(x,Q^2)$
in (\ref{eq:flms}).}
\vspace{-0.4cm}
\label{fig:fl_dyn}
\end{figure}
\begin{figure}[htb]
\vspace{-1.6cm}
\hspace{-0.6cm}
\epsfig{figure= 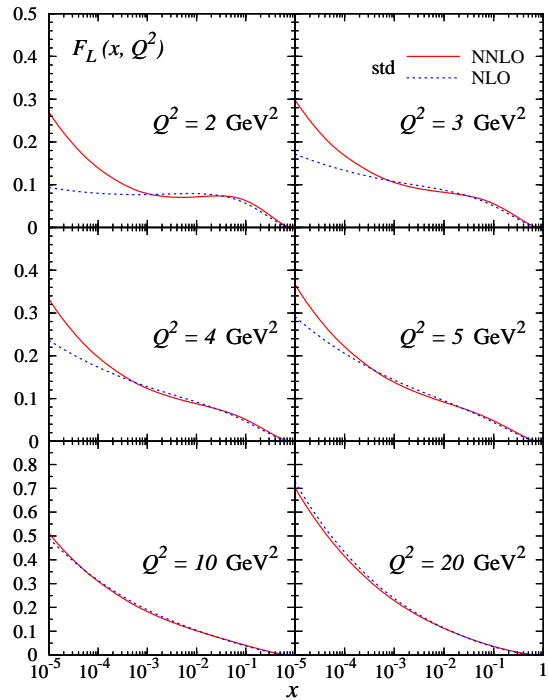, width = 8.cm}
\vspace{-1.5cm}
\caption{As in Figure~\ref{fig:fl_dyn} but for the common standard parton 
distributions.}
\vspace{-0.4cm}
\label{fig:fl_std}
\end{figure}
\begin{figure}[hbt]
\vspace{0.5cm}
\hspace{-0.6cm}
\epsfig{figure= 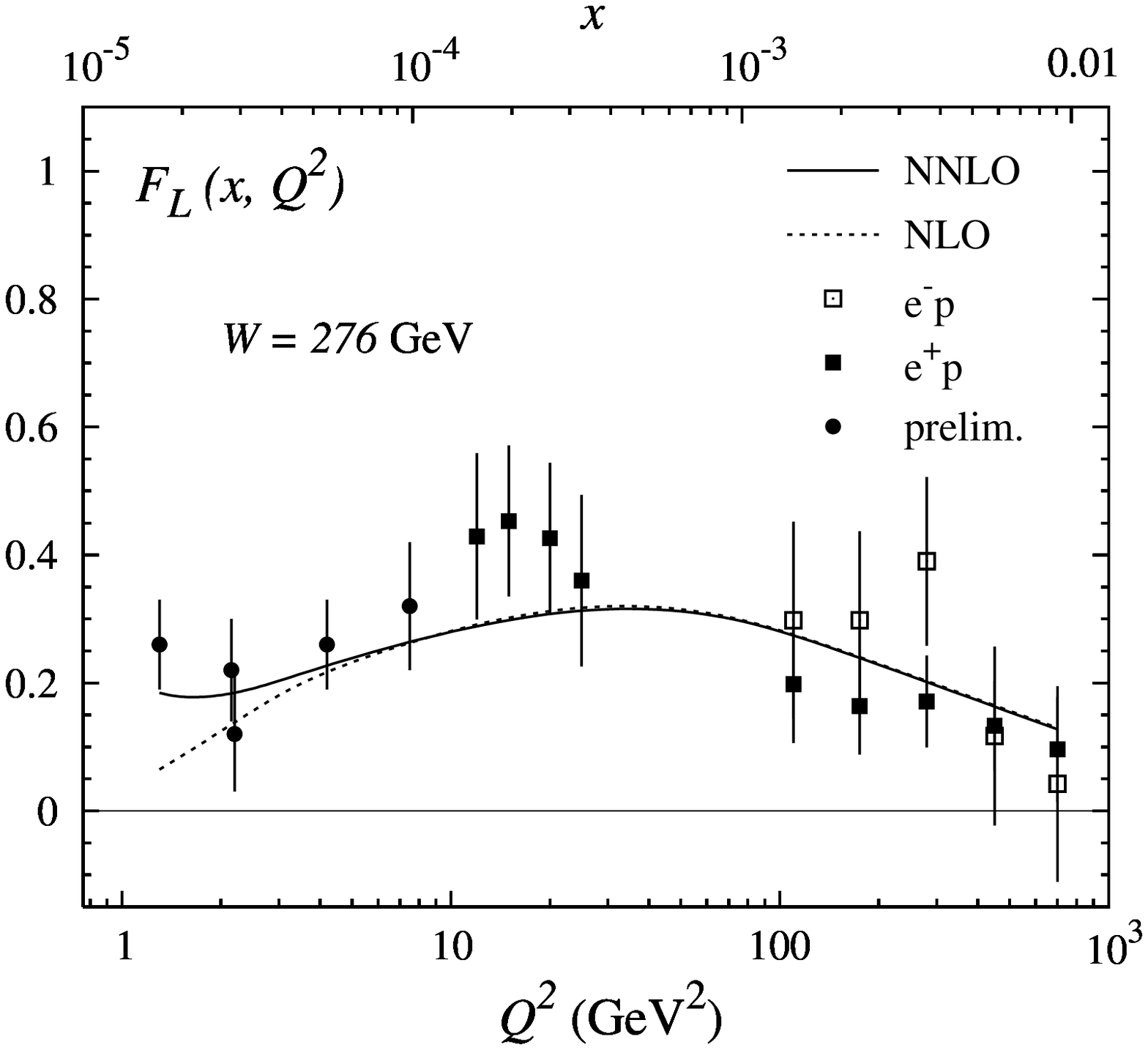, width = 8.2cm}
\caption{Our dynamical NNLO and NLO predictions for $F_L$ at a fixed
value of $W=276$ GeV.  The (partly preliminary) H1 data 
\cite{ref28,ref28b,ref27,ref29}
are at fixed $W\simeq 276$ GeV.}
\label{fig:fl_exp}
\vspace{-0.4cm}
\end{figure}

\section{The longitudinal structure function $F_L(x, Q^2)$}
\label{sec:fl}
In this section  we turn to the perturbative predictions for $F_L(x,Q^2)$.
Similarly to (\ref{eq:f2ms}) one can write for $F_L$
in the $\overline{\rm MS}$ scheme, 
\begin{eqnarray}
x^{-1}F_L & = & C_{L,NS} \otimes q_{NS} \nonumber \\
&&+\frac{2}{9}\, \left( C_{L,q}\otimes q_S +C_{L,g}\otimes g\right) \,
 +x^{-1}F_L^c\nonumber \\
\label{eq:flms}
\end{eqnarray}
where again $\otimes$ in the $n_f=3$ light quark flavor sector denotes the
convolution, $q_{NS}$ stands for the usual flavor non--singlet
combination and $q_S=\sum_{q=u,d,s}(q+\bar{q})$ is the corresponding
flavor--singlet quark distribution.  We use the NLO expression
\cite{ref10,ref11} for $F_L^c$ also in NNLO due to our ignorance of
the ${\cal{O}}(\alpha_s^3)$ NNLO heavy quark corrections. (Notice that
$F_L^c$ is a genuinely subdominant NLO contribution to the total $F_L$,
being less than 10\% even at $Q^2=2$ GeV$^2$ and $x=10^{-5}$ and further
decreases for increasing $x$. Furthermore, the NNLO 3--loop corrections
to $F_L^{c}$ have been calculated recently \cite{ref21} for $Q^2\gg m_c^2$,
but this asymptotic result is neither applicable for our present 
investigation nor relevant for the majority of presently available data
at lower values of $Q^2$.)
The
perturbative expansion of the coefficient functions can be written as
\begin{equation}
C_{L,i}(\alpha_s,x) = \sum_{n=1}\, 
  \left( \frac{\alpha_s(Q^2)}{4\pi}\right)^n \, c_{L,i}^{(n)}(x)\, .
\end{equation}
In LO, $c_{L,ns}^{(1)} =\frac{16}{3}x$, $c_{L,ps}^{(1)}=0$,
$c_{L,g}^{(1)}=24x(1-x)$ and the singlet--quark coefficient function
is decomposed into the non--singlet and a `pure singlet' contribution,
$c_{L,q}^{(n)} = c_{L,ns}^{(n)} + c_{L,ps}^{(n)}$.  Sufficiently
accurate simplified expressions for the exact \cite{ref22,ref23,ref24}
NLO and \cite{ref25} NNLO coefficient
functions $c_{L,i}^{(2)}$ and $c_{L,i}^{(3)}$, respectively, have
been given in \cite{ref1}.  It has been furthermore noted in \cite{ref1}
that especially for $C_{L,g}$ both the NLO and NNLO contributions are
rather large over almost the entire $x$--range.  Most striking,
however, is the behavior of both $C_{L,q}$ and $C_{L,g}$ at very
small values \cite{ref1,ref26} of $x$:  the vanishingly small LO 
parts ($xc_{L,i}^{(1)}\sim x^2$) are negligible as compared to the 
(negative) constant
NLO 2--loop terms, which in turn are completely overwhelmed by the
positive NNLO 3-loop singular corrections $xc_{L,i}^{(3)}\sim -\ln x$.
This latter singular contribution might be indicative for the 
perturbative instability at NNLO \cite{ref1}, as discussed at the 
beginning, but it should be kept in mind that a small--$x$ information
alone is {\em{insufficient}} for reliable estimates of the 
convolutions occurring in (\ref{eq:flms}) when evaluating physical observables.

We display the predictions for the convolutions of the individual light 
$u,d,s$ quark ($F_L^q$) and
gluon ($F_L^g$) contributions in (\ref{eq:flms}) in Figures~\ref{fig:flq} and
 \ref{fig:flg}, 
respectively,
at two characteristic low values of $Q^2$.  (Note that $F_L^q +
F_L^g = F_L-F_L^c$ according to (\ref{eq:flms})).  Although the perturbative
instability of the subdominant quark contribution in Figure \ref{fig:flq} as
obtained in a  standard fit does not improve  for the
dynamical (sea) quark distributions, the instability disappears 
almost entirely for the dominant dynamical gluon contribution already
at $Q^2\simeq 2$ GeV$^2$ as shown in Figure \ref{fig:flg}.  This implies 
that the 
dynamical predictions for the total $F_L(x,Q^2)$ become perturbatively
stable already at the relevant low values of 
$Q^2$ \raisebox{-0.1cm}{$\stackrel{>}{\sim}$} ${\cal{O}}(2-3$ GeV$^2)$
as evident from Figure \ref{fig:fl_dyn}, in contrast to the standard results
 in Figure \ref{fig:fl_std}.  In the latter case the stability has not 
been fully 
reached even at $Q^2 = 5$ GeV$^2$ where the NNLO result at $x=10^{-5}$
is more than 20\% larger than the NLO one.  A similar discrepancy 
prevails for the dynamical predictions in Figure \ref{fig:fl_dyn} at 
$Q^2=2$ GeV$^2$.
This is, however, not too surprising since $Q^2=2$ GeV$^2$ represents
somehow a borderline value for the leading twist--2 contribution to
become dominant at small $x$ values.  This is further corroborated
by the observation that the dynamical NLO twist--2 fit slightly
undershoots the HERA data for $F_2$ at $Q^2 \simeq 2$ GeV$^2$ in the
small--$x$ region (cf.~Figure~1 of \cite{gjr}), which indicates that
nonperturbative (higher twist) contributions to $F_2$ become relevant
for $Q^2$ \raisebox{-0.1cm}{$\stackrel{<}{\sim}$} 2 GeV$^2$ 
\cite{grv98,gjr}.  
The NLO/NNLO instabilities implied by the standard fit results obtained 
in \cite{ref2,ref3} at 
$Q^2$ \raisebox{-0.1cm}{$\stackrel{<}{\sim}$} 5 GeV$^2$ are even
more violent than the ones shown in Figure \ref{fig:fl_std}.  This is mainly 
due to
the negative longitudinal cross section (negative $F_L(x,Q^2)$) 
encountered in \cite{ref2,ref3}. The perturbative stability in any
scenario becomes in general better the larger $Q^2$, typically beyond
5 GeV$^2$ \cite{ref1,ref2,ref3}, as shown in Figures 
\ref{fig:fl_dyn} and \ref{fig:fl_std}.  This is
due to the fact that the $Q^2$--evolutions eventually force any
parton distribution to become sufficiently steep in $x$. 

For 
completeness we finally compare in Figure \ref{fig:fl_exp}
 our dynamical (leading twist)
NNLO and NLO predictions for $F_L(x,Q^2)$ with a representative
selection of (partly preliminary) HERA--H1 data 
\cite{ref28,ref28b,ref27,ref29}.
Our results for $F_L$, being gluon dominated in the small--$x$ region,
are in full agreement with present measurements which is in contrast
to expectations \cite{ref2,ref3} based on negative parton distributions
and structure functions at small values of $x$.  To illustrate the 
manifest positive definiteness of our dynamically generated structure
functions at $Q^2\geq \mu^2 = 0.5$ GeV$^2$ we show $F_L(x,Q^2)$ in
Figure \ref{fig:fl_exp} down to small values of $Q^2$ although leading twist--2
predictions need not necessarily be confronted with data below, say,
2 GeV$^2$. As pointed out in \cite{jr}, where also a study on the 
$\pm 1 \sigma$ uncertainty bands of $F_L$ has been performed, future 
precision measurements of $F_L$ could even distinguish between 
NLO results and NNLO effects in the 
very small--$x$ region.

\section{Summary and conclusions}

To summarize, recent deep inelastic data for the structure function 
$F_2^{p, n}$, without the inclusion of any $F_L$
data, have been analyzed in the 
dynamical and standard parton model approach at  NLO and NNLO of 
perturbative QCD. In both approaches, perturbative QCD evolutions of parton 
distributions
in the (very) small-$x$ region are fully compatible with all 
 high-statistics measurements of the $Q^2$-dependence of $F_2(x,Q^2)$ 
in that region.  
The results turned out to be perturbatively
stable, therefore additional model assumptions concerning further
resummations of subleading small-$x$ logarithms are not required.

Furthermore, the extracted parton distributions have been used to
predict $F_L$. It has been shown 
that the extreme perturbative NNLO/NLO
instability of $F_L$ at low $Q^2$,
noted in \cite{ref2,ref3,ref4}, is an artifact of the commonly utilized  
`standard'
gluon distributions rather than an indication of a genuine problem
of perturbative QCD.  In fact it has been demonstrated that these
extreme instabilities are reduced considerably already at $Q^2 = 2-3$
GeV$^2$ when utilizing the appropriate, dynamically generated, 
parton distributions at NLO and NNLO. 
It is interesting to notice,
once again, the advantage of the dynamical parton model approach to
perturbative QCD.

\section*{ACKNOWLEDGEMENTS}

I wish to thank the organizers for their kind invitation at this very 
interesting workshop. I thank M.~Gl\"uck and E.~Reya for fruitful discussions
and  collaboration on this topic over the past  years. 

This research is part of the 
   research program of the ``Stichting voor Fundamenteel Onderzoek der 
   Materie (FOM)'', which is financially supported by the ``Nederlandse 
   Organisatie voor Wetenschappelijk Onderzoek (NWO)''.


\begin{thebibliography}{60}


\bibitem{ref18} J.~Pumplin et al., CTEQ Collab.,  
                {  JHEP} {07} (2002) 012.

\bibitem{ref2} A.D.~Martin et al., 
               {  Phys.~Lett.} {B 531} (2002) 216.


\bibitem{grv98} M.~Gl\"uck, E.~Reya, A.~Vogt,
               {  Eur.~Phys.~J.} {C 5} (1998) 461.

\bibitem{gjr} M.~Gl\"uck, P.~Jimenez--Delgado, E.~Reya,
               {  Eur.~Phys.~J.} {C 53} (2008) 355.


\bibitem{jr}   P.~Jimenez-Delgado, E.~Reya, arXiv:0810.4274 [hep-ph].

\bibitem{gpr}
  M.~Gl\"uck, C.~Pisano, E.~Reya,
  Phys.\ Rev.\  D {77} (2008) 074002
  [Erratum-ibid.\  D {78} (2008) 019902].

\bibitem{ref8} M.~Gl\"uck, C.~Pisano, E.~Reya, 
               {  Eur.~Phys.~J.} {C 50} (2007) 29.

\bibitem{ref1} S.~Moch, J.A.M.~Vermaseren, A.~Vogt, 
               {  Phys.~Lett.} 
               {B 606} (2005) 123, and references therein.
\bibitem{ref3} A.D.~Martin, W.J.~Stirling, R.S.~Thorne, 
               {  Phys.~Lett.}
               {  B 635} (2006) 305.
\bibitem{ref4} R.S.~Thorne, Proceedings of the Ringberg Workshop on
               `New Trends in HERA Physics' (Tegernsee, Oct.~2005), 
               p.\ 359 (hep--ph/0511351).

\bibitem{ref7} C.D.~White, R.S.~Thorne, 
               {  Phys.~Rev.} {  D 75} (2007) 034005.       




\bibitem{grheavy}
  M.~Gl\"uck, E.~Reya,
  Mod.\ Phys.\ Lett.\  A {22} (2007) 351.

\bibitem{curvref3} W.L.~van Neerven, A.~Vogt, {  Nucl.~Phys.}
               {B 568} (2000) 263. 
\bibitem{curvref4} W.L.~van Neerven, A.~Vogt, {  Nucl.~Phys.}
               {B 588} (2000) 345 and arXiv:hep-ph/0006154
               (corrected).

\bibitem{curvref5} J.~Bl\"umlein, A.~Vogt,
               {  Phys.~Rev.} {D 58} (1998) 014020.
\bibitem{curvref6} W.~Furmanski, R.~Petronzio, {  Z.~Phys.} 
               {C 11} (1982) 293, and references therein. 
\bibitem{curvref7} S.~Moch, J.A.M.~Vermaseren, A.~Vogt, 
               {  Nucl.~Phys.} 
               {B 688} (2004) 101.  

\bibitem{ref16} M.~Gl\"uck, E.~Reya, C.~Schuck,
                {  Nucl.~Phys.} 
                {  B 754} (2006) 178.


\bibitem{curvref9} A.~Vogt, S.~Moch, J.A.M.~Vermaseren, 
               {  Nucl.~Phys.} {B 691} (2004) 129.

\bibitem{ref9} A.~Vogt, {  Comput.~Phys.~Commun.} {170}
               (2005) 65. 


\bibitem{curvref2} M.~Gl\"uck, C.~Pisano, E.~Reya, 
               {  Eur.~Phys.~J.} {C 40} (2005) 515.

\bibitem{ref10} E.~Laenen, S.~Riemersma, J.~Smith, W.L.~van Neerven,
               {  Nucl.~Phys.} {B 392} (1993) 162.
\bibitem{ref11} S.~Riemersma, J.~Smith, W.L.~van Neerven, 
                {  Phys.~Lett.}
                {B 347} (1995) 143. 

\bibitem{Gluck:1993dpa}  M.~Gl\"uck, E.~Reya, M.~Stratmann,
  Nucl.\ Phys.\  B {422} (1994) 37.


\bibitem{Gluck:1994uf}
  M.~Gl\"uck, E.~Reya, A.~Vogt, Z.\ Phys.\  C {67} (1995) 433.

\bibitem{Vogt:1996wr} 
  A.~Vogt, DESY 96-012, Proc. of DIS '96, Rome, April 1996, ed. 
  by G. D'Agostini, A. Nigro (World Scientific, 1997) p. 254,   
  hep-ph/9601352.


\bibitem{ref12} S.I.~Alekhin, 
                {  Phys.~Rev.}
                {  D 68} (2003) 014002.
\bibitem{ref13} S.I.~Alekhin,
                {  JETP Lett.} {  82} (2005) 628.
\bibitem{ref14} S.I.~Alekhin, K.~Melnikov, F.~Petriello,
                {  Phys.~Rev.}
                {D 74} (2006) 054033. 


\bibitem{ref28} C.~Adloff et al., H1 Collab., 
                {  Eur.~Phys.~J.} {C 21} (2001) 33.

\bibitem{ref28b} C.~Adloff et al., H1 Collab., 
                {  Eur.~Phys.~J.} {C 30} (2003) 1.

\bibitem{ref28c} A.C.~Benvenuti et al., BCDMS Collab., 
                {  Phys.~Lett.} {B 223} (1989) 485;
                {B 237} (1990) 599.

\bibitem{ref28d} M.~Arneodo et al., NMC Collab., {Nucl.~Phys.} 
                {B 483} (1997) 3; {B 487} (1997) 3. 


\bibitem{ref15} A.L.~Kataev et al.,
                {Phys.~Lett.} {B 388} (1996) 179;
                {B 417} (1998) 374.


\bibitem{ref17} J.~Bl\"umlein, H.~B\"ottcher, A.~Guffanti,
                {  Nucl.~Phys.} {B} (Proc.~Suppl.) {135} 
                (2004) 152;
                {  Nucl.~Phys.} {B 774} (2007) 182.

\bibitem{ref19} W.K.~Tung et al., CTEQ Collab.,
                {  JHEP} {  02} (2007) 053.

\bibitem{ref20} J.~Bl\"umlein, DIS 2007 (Munich, April 2007),
                arXiv:0706.2430.


\bibitem{curvref19} S.~Forte, G.~Altarelli, R.D.~Ball, talk presented
                at {DIS 2006}, Tsukuba, Japan (April 2006),
                and references therein (arXiv:hep-ph/0606323).




\bibitem{ref21} J.~Bl\"umlein et al., {Nucl.~Phys.} {  B 755}
                (2006) 272.

\bibitem{ref22} D.I.~Kazakov, A.V.~Kotikov, {Nucl.~Phys.} {  B 307}
                (1988) 721  [Erratum-ibid.\  B {345} (1990) 299].

\bibitem{ref23} J.~Sanchez Guillen et al., {Nucl.~Phys.}
                {  B 353} (1991) 337.

\bibitem{ref24} E.B.~Zijlstra, W.L.~van Neerven, {Phys.~Lett.}
                {B 273} (1991) 476.

\bibitem{ref25} J.A.M.~Vermaseren, A.~Vogt, S.~Moch, {Nucl.~Phys.}
                {B 724} (2005) 3.

\bibitem{ref26} S.~Catani, F.~Hautmann, {Nucl.~Phys.} {B 427}
                (1994) 475.

\bibitem{ref27} C.~Adloff et al., H1 Collab., {Phys.~Lett.} {B 393}
                (1997) 452.


\bibitem{ref29} E.M.~Lobodzinska, H1 Collab., DIS 2004
                (Strbske Pleso, Slovakia),
                hep--ph/0311180;\\
                T.~Lastovicka, H1 Collab., 
                {Eur.~Phys.~J.} {C 33} (2004) s388.

\end{thebibliography}
\end{document}